\documentclass[letter,aps,showpacs,superscriptaddress,nofootinbib,tightenlines,prl]{revtex4}

\usepackage{amsfonts}
\usepackage{multirow}
\usepackage{mathrsfs}
\usepackage{graphicx}
\usepackage{amsmath}
\usepackage{amssymb}
\usepackage{bm}
\usepackage{bbm}
\usepackage{color}
\usepackage{hyperref}

\def\OMIT#1{}

\newcommand{\nn}{\nonumber}

\newcommand{\beq}{\begin{equation}}
\newcommand{\eeq}{\end{equation}}
\newcommand{\bqa}{\begin{eqnarray}}
\newcommand{\eqa}{\end{eqnarray}}

\newcommand{\bseq}{\begin{subequations}}
\newcommand{\eseq}{\end{subequations}}

\begin{document}


\title{\mbox{}\\[14pt]
Fragmentation production of fully-charmed tetraquarks at LHC
}

\author{Feng Feng~\footnote{F.Feng@outlook.com}}
\affiliation{Institute of High Energy Physics, Chinese Academy of
	Sciences, Beijing 100049, China\vspace{0.2 cm}}
\affiliation{China University of Mining and Technology, Beijing 100083, China\vspace{0.2 cm}}

\author{Yingsheng Huang~\footnote{huangys@ihep.ac.cn}}
\affiliation{Institute of High Energy Physics, Chinese Academy of
	Sciences, Beijing 100049, China\vspace{0.2 cm}}
\affiliation{School of Physics, University of Chinese Academy of Sciences,
	Beijing 100049, China\vspace{0.2 cm}}

\author{Yu Jia~\footnote{jiay@ihep.ac.cn}}
\affiliation{Institute of High Energy Physics, Chinese Academy of
Sciences, Beijing 100049, China\vspace{0.2 cm}}
\affiliation{School of Physics, University of Chinese Academy of Sciences,
Beijing 100049, China\vspace{0.2 cm}}

\author{Wen-Long Sang~\footnote{wlsang@ihep.ac.cn}}
\affiliation{School of Physical Science and Technology, Southwest University, Chongqing 400700, P.R. China}

\author{Xiaonu Xiong~\footnote{xnxiong@csu.edu.cn}}
\affiliation{School of Physics and Electronics, Central South University, Changsha 418003, China}

\author{Jia-Yue Zhang~\footnote{zhangjiayue@ihep.ac.cn}}

\affiliation{Institute of High Energy Physics, Chinese Academy of Sciences, Beijing 100049, China\vspace{0.2 cm}}
\affiliation{School of Physics, University of Chinese Academy of Sciences,
	Beijing 100049, China\vspace{0.2 cm}}

\date{\today}

\begin{abstract}
The $X(6900)$ resonance,  very recently discovered in the double-$J/\psi$ channel at LHCb experiment,
has spurred intensive interest in unravelling the nature of the fully charmed tetraquark state.
The aim of this paper is to present a model-independent theoretical framework to study the inclusive production of
this novel species of exotic hadrons, the resonances composed of four heavy quark (commonly referred to as $T_{4c}$),
at large $p_T$ in hadron collision experiments.
Appealing to asymptotic freedom and the fact $m_c\gg \Lambda_{\rm QCD}$,
we propose that the nonpertubative yet universal gluon-to-$T_{4c}$ fragmentation function,
can be decomposed into the product of the perturbatively calculable short-distance coefficient and the
long-distance NRQCD matrix elements. We compute the short-distance coefficient at lowest-order in $\alpha_s$ and velocity
expansion. Adopting the diquark ansatz to roughly estimate those not-yet-known NRQCD matrix elements, together with
the standard QCD factorization theorem, we predict the differential production rates for the
$T_{4c/4b}(0^{++})$ and $T_{4c/4b}(2^{++})$  at large $p_T$ in $pp$ collision,
which eagerly awaits the confrontation with the future LHC experiments.
\end{abstract}

\maketitle

\paragraph{\color{blue}Introduction}
Since the discovery of $X(3872)$ in 2003, many exotic charmonium-like states have been discovered at {\tt B} factories,
{\tt LHC} and {BESIII} experiments, collectively referred to as $XYZ$ states (for recent reviews, see \cite{Liu:2019zoy,Chen:2016qju,Klempt:2007cp,Hosaka:2016pey,Ali:2017jda}). Most of the $XYZ$ states do not
seem to fit into the conventional $c\bar{c}$ category in quark model.
The debate is still actively ongoing, whether some of the $XYZ$ states might be identified with the
compact tetraquark, or loosely-bound molecule, or even merely originate from kinematic effects.
Unfortunately, since the light quark/antiquarks can readily pop out of QCD vacuum,
many interpretations of $XYZ$ as the unconventional multiquark states,
which are mainly coined in the context of phenomenological quark model,
have often been criticized to have rather vague connection to QCD.
Needless to say, it is the ultimate goal of hadron physics to firmly understand the nature of the $XYZ$ states
directly from the first principle of QCD.

Very recently, {\tt LHCb} collaboration has announced the discovery of a narrow structure near $6.9$ GeV in the di-$J/\psi$ invariant mass spectrum, with a global significance of more than $5\sigma$~\cite{Aaij:2020fnh}.
Some additional broad structures ranging from 6.2 to 6.8 GeV are also reported. The narrow resonance near 6.9 GeV,
dubbed as $X(6900)$, is widely believed to be a strong candidate for the fully-charmed tetraquark.
The fully charmed tetraquarks (hereafter labeled by $T_{4c}$) open a unique and novel window for the study of
exotic hadrons. Since the charm quark is too heavy to be readily excited from the vacuum,
the charm/anticharm quark number inside $X(6900)$ may be viewed as the separately conserved
quantum numbers, therefore the Fock component $|c\bar{c}c\bar{c}\rangle$ appears to be an
unambiguous and meaningful assignment.

The existence of fully heavy tetraquarks was originally envisaged four decades ago~\cite{Iwasaki:1976cn,Chao:1980dv,Ader:1981db}.
Recently a flurry of work have reemerged along this direction.
The mass spectra and decay patterns of $T_{4c}$ have been studied in quark potential model~\cite{Wu:2016vtq,Barnea:2006sd,Badalian:1985es,Liu:2019zuc,Bedolla:2019zwg,Lloyd:2003yc,
Anwar:2017toa,Esposito:2018cwh,Bai:2016int,Richard:2017vry,
Berezhnoy:2011xn,Becchi:2020uvq,Heller:1985cb,Debastiani:2017msn,Lu:2020cns,liu:2020eha,Yang:2020rih},
as well as QCD sum rules~\cite{Chen:2016jxd,Chen:2018cqz,Wang:2017jtz,Wang:2018poa,Chen:2020xwe,Wang:2020ols}.
It is worth remarking that, lattice NRQCD has also attempted to investigate the lowest-lying spectrum of the
$bb\bar b\bar b$ sector, but found no indication of any states below $2\eta_b$ threshold in the $0^{++}$, $1^{+-}$ and $2^{++}$ channel~\cite{Hughes:2017xie}.
Since the mass of $X(6900)$ is above the double $J/\psi$ threshold, it is unlikely to be identified with the molecule formed by two charmonia.
Chen {\it et al.} adopted QCD sum rule to interpret that the broad structure can be interpreted as an $0^{++}$ tetraquark state while the narrow as a $P$-wave\cite{Chen:2020xwe}.  The $X(6900)$ is also interested  as the radical excitation of $0^{++}$~\cite{Wang:2020ols,Karliner:2020dta,Lu:2020cns} .
On the other hand, the $T_{4c}$ family are widely regarded as the compact tetraquark.
In particular, the most popular phenomenological picture is to portray the $T_{4c}$ as the color-singlet state formed by
the diquark-antidiquark cluster.

While the spectra and decay properties are studied at length, the production mechanism of fully heavy tetraquark was rarely mentioned, even for highly model-dependent ones\cite{Karliner:2016zzc,Berezhnoy:2011xy,Berezhnoy:2011xn,Becchi:2020mjz,Becchi:2020uvq,Maciula:2020wri,Carvalho:2015nqf}. A popular method is to employ so-called quark-hadron duality between parton and hadron cross sections, in order to obtain not the cross section of tetraquark production, but rather the resonant cross section of the quarkonium pair as the decay product of the tetraquark\cite{Berezhnoy:2011xy,Berezhnoy:2011xn}. A rough estimate of the tetraquark cross section (0.7 nb at $\sqrt{s}=7$ TeV) was obtained by comparing the resonant and nonresonant double quarkonia production\cite{Karliner:2016zzc}. Similar ideas of transfroming tetraquark cross section into known quarkonium cross sections can also be found with \cite{Becchi:2020mjz,Becchi:2020uvq}. Another approach is to treat the hadronisation process with (modified) color evaporation model. Carvalho et al. calculated $T_{4c}$ production for double parton scattering mechanism at LHC, with a relation to X(3872) production cross section obtained from so-called ``pocket" formula\cite{Carvalho:2015nqf}. They arrived with results of $3.6\pm2.5$ nb at $\sqrt{s}=7$ TeV and $7.0\pm4.8$ nb at $\sqrt{s}=14$ TeV. Later Maciuła et al. studied both single parton scttering and double parton scattering mechanisms of tetraquark production with color evaporation model as well, but instead they calculated the resonance production of $T_{4c}$ with $J/\psi$ pair production\cite{Maciula:2020wri}.

\paragraph{\color{blue} QCD Factorization theorem for high-$p_T$ production of $T_{4c}$ \label{QCD:fact:theor}}
Our central goal is to present a model-independent framework to describe inclusive production at {\tt LHC}.
To make life simpler, we concentrate on the high-$p_T$ production of $T_{4c}$, {\it i.e.}, in the limit $p_T\gg 4m_c$.
The celebrated QCD factorization theorem~\cite{Collins:1989gx} indicates
that the dominant inclusive production of a high-$p_T$ hadron
is via the fragmentation mechanism,
and the differential cross section of $T_{4c}$ in hard-scattering $pp$ collision can be expressed as the following factorized form:
\bqa
		\mathrm{d} \sigma\left(p p \rightarrow T_{4c}\left(p_{\mathrm{T}}\right)+X\right) &=&\sum_{i}\int_0^1 \mathrm{d} x_a  \int_0^1 \mathrm{d} x_b \int_{0}^{1} \mathrm{d} z\; f_{a/p}(x_a,\mu)f_{b/p}(x_b,\mu) \nn\\
		&\times & d {\hat \sigma}(a b \rightarrow i(p_T/z)+X, \mu)  D_{i \rightarrow T_{4c}}\left(z,\mu\right)+{\cal O}(1/p_T),
\label{QCD:fact:theor}
\eqa
where $f_{a,b/p}$ denotes the parton distribution functions (PDFs) of a proton, $d {\hat \sigma}$ represents the
partonic cross section, and $D_{i \rightarrow T_{4c}}$ designates the fragmentation function for parton $i$ into
$T_{4c}$, and $z\in [0,1]$ is the ratio of the light-cone momentum carried by $H$ with respect to that by
the parent parton $i$.

For fully heavy tetraquark production at {\tt LHC}, the partonic channel $gg\to gg$ is much more important
than  $gg\to q\bar{q}$. Therefore, for simplicity,
we will be content with focusing only on gluon-to-$T_{4c}$ fragmentation.

\paragraph{\color{blue} Gluon-to-$T_{4c}$ fragmentation function from Collins-Soper definition\label{FF:NRQCD:factorization}}
The fragmentation function is a nonperturbative yet process-independent object, which accommodates
a rigorous field-theoretic definition given by Collins and Soper in 1981~\cite{Collins:1981uw}.
The gluon-to-$T_{4c}$ fragmentation function is defined by
\begin{align}
	\begin{aligned}[b]
		D_{g \rightarrow T_{4c}}(z, \mu)= &\frac{-g_{\mu \nu} z^{d-3}}{2 \pi k^{+}\left(N_{c}^{2}-1\right)(d-2)} \int_{-\infty}^{+\infty} d x^{-} e^{-i k^{+} x^{-}}
\\
		& \times\sum_X \left\langle 0\left|G_{c}^{+\mu}(0) \mathcal{E}^{\dagger}\left(0,0, \mathbf{0}_{\perp}\right)_{c b}
|T_{4c}(P)+X\rangle\langle T_{4c}(P)+X| \mathcal{E}\left(0, x^{-}, \mathbf{0}_{\perp}\right)_{b a} G_{a}^{+\nu}\left(0, x^{-}, \mathbf{0}_{\perp}\right)\right| 0\right\rangle,
\end{aligned}
\label{CS:def:gluon:T4c:frag}
\end{align}
with spacetime dimension $d=4-2\epsilon$. $G^{\mu\nu}$ denotes the gluonic field-strength tensor,
with $k$ representing the momentum inserted by this strength tensor, and $P$ represents the momentum carried by $T_{4c}$.
$z$ is the light-cone momentum fraction $z = P^+/k^+$, and $\mu$ signifies the renormalization scale.
The sum in \eqref{CS:def:gluon:T4c:frag} implies that all the asymptotic states
containing $T_{4c}$ and any additional soft hadrons should be included, meanwhile the polarization of $T_{4c}$
should be summed. $\mathcal{E}$ is the gauge link in $SU(3)$ adjoint representation to ensure manifest gauge invariance.

The fragmentation function $D_{g \rightarrow T_{4c}}(z, \mu)$ obeys the renowned
Dokshitzer-Gribov-Lipatov-Altarelli-Parisi (DGLAP) evolution equation:
\beq
\mu \frac{\partial}{\partial \mu} D_{g \rightarrow T_{4c}}(z, \mu)= \sum_{i\in\{g, c\}} \int_{z}^{1} \frac{\mathrm{d} y}{y} P_{g\rightarrow i}\left(\frac{z}{y}, \mu\right) D_{i \rightarrow T_{4c}}(y, \mu).
\label{DGLAP:evolu:eq}
\eeq
For concreteness, the gluon-to-gluon splitting kernel at lowest order reads
\beq
P_{g\rightarrow g}(z,\mu)=\frac{6\alpha_s(\mu)}{\pi}\left[\frac{(1-z)}{z}+\frac{z}{(1-z)_{+}}+z(1-z)
	+\left(\frac{11}{12}-\frac{n_{f}}{18}\right) \delta(1-z)\right],
\eeq
where $n_f$ denotes the number of active light quarks.

\paragraph{\color{blue} NRQCD factorization for gluon-to-$T_{4c}$ fragmentation \label{NRQCD:factorization:FF}}
Undoubtedly, the fragmentation functions for light hadrons are genuinely nonperturbative objects,
which can only be extracted via experimental measurements.
However, the fragmentation function for heavy hadrons, exemplified by heavy quarkonium,
is a dramatically different story.
Prior to hadronization, the heavy quark and antiquark have to be created at rather short distance $\sim 1/m$,
therefore it is plausible to invoke asymptotic freedom to separate the parton fragmentation into quarkonium into the product of
the perturbatively calculable short-distance coefficients (SDC) and nonperturbative long-distance matrix elements (LDME).
The systematization of this line of argument leads to the influential NRQCD factorization approach,
according to which the parton fragmentation into quarkonium $H$ can be expressed as
\beq
D_{g \rightarrow H}(z)=\sum_{n} d_{n}(z)\left\langle 0\left|\mathcal{O}_{n}^{H}\right| 0\right\rangle,
\label{NRQCD:fac:quarkonium:frag}
\eeq
with $d_n(z)$ denoting the SDCs, and $\mathcal{O}_{n}^{H}$ denoting various vacuum matrix elements of NRQCD production operators.
A double expansion in quark velocity $v$ and $\alpha_s$ is implicitly
understood in \eqref{NRQCD:fac:quarkonium:frag}.
In the past two decades, a handful of fragmentation functions for a variety of
quarkonia have been analyzed in this framework~\cite{Braaten:1994xb,Ma:1994zt,Ma:2013yla}.

One may naturally ask whether the NRQCD factorization formula \eqref{NRQCD:fac:quarkonium:frag} is also applicable
for fragmentation function for $T_{4c}$? A key observation is that in order to produce a $T_{4c}$ state,
prior to hadronization, two $c$ and two $\bar{c}$ have to be first created at short distance $\sim 1/m$,
thus one can still invoke asymptotic freedom to separate the fragmentation function into a short-distance and long-distance part.
This strongly suggest the NRQCD factorization ansatz should hold for tetraquark case.
In accordance with \eqref{NRQCD:fac:quarkonium:frag}, the fragmentation function
for gluon into the $0(2)^{++}$ tetraquark can be expressed as
 \begin{align}
\notag D_{g \rightarrow T_{4 c}}\left(z, \mu_{\Lambda}\right)=&\frac{d_{3, 3}\left[g \rightarrow c c \bar{c} \bar{c}^{(J)}\right]}{m^{9}}\left|\left\langle 0\left|\mathcal{O}_{\mathbf{3}\otimes\mathbf{\bar 3}}^{(J)}\right| T_{4 c}^{(J)}\right\rangle\right|^{2}+\frac{d_{6,6}\left[g \rightarrow c c \bar{c} \bar{c}^{(J)}\right]}{m^{9}}\left|\left\langle 0\left|\mathcal{O}_{\mathbf{6} \otimes \bar{\mathbf{6}}}^{(J)}\right| T_{4 c}^{(J)}\right\rangle\right|^{2}\\
 &+\frac{d_{3, 6}\left[g \rightarrow c c \bar{c} \bar{c}^{(J)}\right]}{m^{9}} 2{\rm Re}\left[\left\langle 0\left|\mathcal{O}_{\mathbf{3}\otimes\mathbf{\bar 3}}^{(J)}\right| T_{4 c}^{(J)}\right\rangle\left\langle  T_{4 c}^{(J)}\left|\mathcal{O}_{\mathbf{6} \otimes \bar{\mathbf{6}}}^{(J)\dagger} \right|0\right\rangle\right]+\cdots,
 \label{NRQCD:fac:Tc:fragmentation}
\end{align}
where the prefactor $1/m^9$ is deliberately chosen to ensure that the SDCs
$d_n(z)$ are dimensionless.
In \eqref{NRQCD:fac:Tc:fragmentation} we have retained only the lowest-order terms in velocity expansion,
and also utilized the vacuum-saturation approximation to enhance the predictive capability.
$\mathcal{O}^{(J)}_{\rm color}$ ($J=0,2$) represent the
composite color-singlet operators comprising four NRQCD fields,
which possesses the same quantum number with the physical $0(2)^{++}$ tetraquarks:
\begin{subequations}
\begin{align}
 &\mathcal{O}^{(0)}_{\mathbf{3}\otimes\mathbf{\bar 3}}=-\frac{1}{\sqrt{3}}[\psi_a^T(i\sigma^2)\sigma^i\psi_b] [\chi_c^{\dagger}\sigma^i (i\sigma^2)\chi_d^*]\;
 \mathcal{C}^{ab;cd}_{\mathbf{3}\otimes\bar{\mathbf{3}}},
\\
 &\mathcal{O}^{\alpha\beta;(2)}_{\mathbf{3}\otimes\mathbf{\bar 3}}=\frac{1}{2}[\psi_a^T(i\sigma^2)\sigma^m\psi_b] [\chi_c^{\dagger}\sigma^n(i\sigma^2)\chi_d^*]\;\Gamma^{\alpha\beta;mn}
 \;\mathcal{C}^{ab;cd}_{\mathbf{3}\otimes\bar{\mathbf{3}}},
\\
&\mathcal{O}^{(0)}_{\mathbf{6}\otimes\bar{\mathbf{6}}}=\frac{1}{\sqrt{6}}
[\psi_a^T(i\sigma^2)\psi_b] [\chi_c^{\dagger}(i\sigma^2)\chi_d^*]\;
\mathcal{C}^{ab;cd}_{\mathbf{6}\otimes\bar{\mathbf{6}}},
\end{align}
\label{NRQCD:composite:operators}
\end{subequations}
where $\sigma^i$ are Pauli matrices, $\psi$ and $\chi^\dagger$ are the standard NRQCD fields annihilating the $c$ and $\bar c$, respectively.
The rank-4 Lorentz tensor is given by $\Gamma^{\alpha\beta;mn}\equiv g^{\alpha m} g^{\beta n}+g^{\alpha n} g^{\beta m}-\frac{1}{2} g^{\alpha \beta} g^{mn}$,
and the rank-4 color tensors read
\begin{subequations}
\bqa
&& \mathcal{C}^{ab;cd}_{\mathbf{3} \otimes\bar{\mathbf{3}}}\equiv \frac{1}{(\sqrt{2})^2} \epsilon^{abm}\epsilon^{cdn}\frac{\delta^{mn}}{\sqrt{3}}=\frac{1}{2\sqrt{N_c}}(\delta^{ac}\delta^{bd}-\delta^{ad}\delta^{bc})
\\
&& \mathcal{C}^{ab;cd}_{\bar{\mathbf{6}}\otimes\mathbf{6}}
 \equiv \frac{1}{2\sqrt{6}}(\delta^{ac}\delta^{bd}+\delta^{ad}\delta^{bc}).
\eqa
\label{color:tensor}
\end{subequations}
Note that these NRQCD operators in \eqref{NRQCD:composite:operators} can also be inferred by performing the
FWT transformation from the QCD interpolating currents given in \cite{Chen:2020xwe}.

For convenience, we have deliberately chosen the composite operators in
\eqref{NRQCD:composite:operators} in a diquark-antidiquark format. Therefore,
to ensure the overall color neutrality,
the color associated with the diquark and anti-diquark operators
can be decomposed either as $\mathbf{3}\otimes\mathbf{\bar 3}$ or
$\mathbf{6}\otimes\mathbf{\bar 6}$.
We emphasize the operators given in \eqref{NRQCD:composite:operators} span
the general dimension-6 operator basis (without derivative), and one can employ Fierz identity to rearrange
the operators as molecular-like form with the color structure
$\mathbf{1}\otimes\mathbf{1}$ or as $\mathbf{8}\otimes\mathbf{8}$~\footnote{Note the operator $\mathcal{O}^{(2)}_{\mathbf{6}\otimes\bar{\mathbf{6}}}$ is absent in
\eqref{NRQCD:composite:operators}, because such operator is incompatible with the
Fermi statistics in constructing the $\bar{\mathbf{6}}\otimes\mathbf{6}$ diquark-antidiquark cluster
for $2^{++}$ tetraquark.  Consequently, the $d_{6,6}$ term and the interference term $d_{3,6}$
in \eqref{NRQCD:fac:Tc:fragmentation} are absent for $2^{++}$ tetraquark.}.

To proceed, we employ the familiar perturbative matching method to determine the three SDCs in
\eqref{NRQCD:fac:Tc:fragmentation}.
The punchline is that SDCs are insensitive to the long-distance binding mechanism of a tetraquark,
thus it is legitimate to replace $T_c$ by a fictitious ``tetraquark'' state $\vert [cc][\bar{c}\bar{c}]\rangle$ in \eqref{NRQCD:fac:Tc:fragmentation}, which are formed by four free charm quarks. To expedite the perturbative
 matching procedure, we construct the fictitious tetraquark as the free diquark-antidiquark pair. The overall color neutrality require the color structure of diquark-antidiquark pair must be either $\mathbf{3}\otimes\mathbf{\bar 3}$ or $\mathbf{6}\otimes\mathbf{\bar 6}$. If we only
 consider the simplest $S$-wave diquark, Fermi statistics then enforces that the $[cc]_{\mathbf{\bar 3}}$ diquark
 must form a spin triplet, while $[cc]_{\mathbf{6}}$ diquark must form a spin singlet.
 If we assume the orbital angular momentum between diquark and antidiquark cluster is again $S$ wave,
 the tetraquark could carry the $J^{PC}$ quantum number of
 $0^{++}$, $1^{+-}$ or $2^{++}$ for the color configuration $\mathbf{3}\otimes\mathbf{\bar 3}$, while carry
the quantum number only of  $0^{++}$ for the color configuration $\mathbf{6}\otimes\mathbf{\bar 6}$~\footnote{In literature, since
the inter-quark color force is attractive for $[cc]_{\mathbf{\bar 3}}$, and repulsive for
$[cc]_{\mathbf{6}}$, the latter is often referred to as ``bad'' diquark, and discarded in potential model analysis.
From perspective of NRQCD factorization, there is no convincing reason to exclude the
contribution due to $\mathcal{O}^{(0)}_{\mathbf{6}\otimes\bar{\mathbf{6}}}$, so we retain the
$\mathbf{6}\otimes\mathbf{\bar 6}$ constitute in the Fock state of a tetraquark.}.

\begin{figure}[hbtp]
\centering
\includegraphics[width=0.3\textwidth]{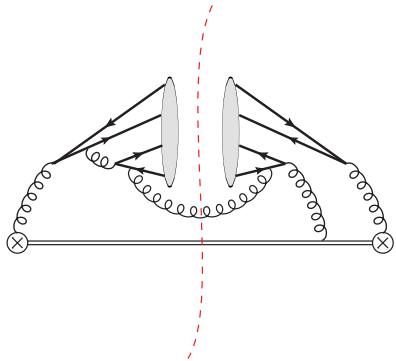}
\caption{A representative Feynman diagram for the fragmentation function of gluon into $T_{4c}$. The grey blob indicates
the $C$-even tetraquark. Horizontal double line denotes the eikonal line.}
\label{tetra-diagrams}
\end{figure}

Our task is then to compute both sides of \eqref{NRQCD:fac:Tc:fragmentation} in perturbative QCD and NRQCD, to solve for the desired SDCs.
The calculation in the pNRQCD side is very straightforward.
For the pQCD computation, we follow the Collins-Soper definition in \eqref{CS:def:gluon:T4c:frag} by replacing the $T_c$ with the fictitious
tetraquark. Since {\tt LHCb} observes the possible tetraquark in the double-$J/\psi$ spectrum, we concentrate on the $C$-even
tetraquark only. Therefore, we choose not to consider the $1^{+-}$ state.
We work in Feynman gauge, and use the dimensional regularization to regularize the potential UV and IR divergences.
To expedite the calculation, we employ the self-written program {\tt HepLib}, which employ {\tt Qgraf}~\cite{Nogueira:1991ex} and {\tt GiNaC}~\cite{Bauer:2000cp} internally to generate the Feynman diagrams and amplitudes for the partonic fragmentation function $D_{g\to cc\bar{c}\bar{c}}$. At the lowest order in $\alpha_s$, there are 98 diagrams for each side of the cut, with one typical diagram displayed
in Fig.~\ref{tetra-diagrams}. Note it is the $C$-parity conservation that demands an additional gluon to be emitted from $c$ or $\bar{c}$ and
across the cut line. To the lowest order in $v$, we set two $c$ and two $\bar{c}$ to share equal momentum. Furthermore, we employ the following
projector in the amplitude to enforce $|[cc] [bar{c}\bar{c}]$ to be in the desired spin/color quantum number
\beq
	\bar u^a_i\bar u^b_j v^c_k v^d_l\to (\textsf{C}\Pi_\mu)^{ij}(\Pi_\nu \textsf{C} )^{lk}\mathcal{C}^{ab;cd}_{\rm color} J^{\mu\nu}_{0,1,2},
\eeq
where $\textsf{C}=i\gamma^2\gamma^0$ is the charge conjugation matrix, the color tensor $\mathcal{C}^{ab;cd}$
is given in \eqref{color:tensor}, $\Pi_\mu$ is the spin-triplet projector of two fermion\cite{Petrelli:1997ge}, and
$J^{\mu\nu}_{0,1,2}$ are the spin projectors of diquark/anti-diquark cluster, quite similar in structure
to the orbital angular momentum projection in \cite{Braaten:2002fi}.

After performing partial fraction, we end up with very simple master integrals containing only one propagator, which can be straightforwardly calculated. We also make a crosscheck by using {\tt FeynArts}~\cite{Hahn:2000kx} and {\tt FeynCalc}~\cite{Shtabovenko:2016sxi} to recompute the
process and find agreement. by solving equation in \eqref{NRQCD:fac:Tc:fragmentation}, we finally obtain the intended SDCs $d_n(z)$.
Reassuringly, these SDCs are free of IR divergences, despite the fact that individual Feynman diagrams may be IR divergent.
This is as expected from color transparency consideration, since the soft gluon cannot resolve the details a composite $S$-wave
color-singlet object.

For gluon fragmentation into $0^{++}$ tetraquark, the three SDCs listed
in \eqref{NRQCD:fac:Tc:fragmentation} read
\bseq
\begin{align}
\notag d_{3, 3}\left(g \rightarrow  0^{++}\right)=& \frac{\pi^{2} \alpha_{s}^{4}}{497664 z(2-z)^{2}(3-z)}\left[186624-430272 z+511072 z^{2}-425814 z^{3}\right.\\
\notag &+217337 z^{4}-61915 z^{5}+7466 z^{6}+42(1-z)(2-z)(3-z)(-144+634 z\\
 &\left.-385 z^{2}+70 z^{3}\right) \log (1-z)+36(2-z)(3-z)\left(144-634 \notag z+749 z^{2}-364 z^{3}\right.\\
\notag  &\left.+74 z^{4}\right) \log \left(1-\frac{z}{2}\right)+12(2-z)(3-z)\left(72-362 z+361 z^{2}-136 z^{3}+23 z^{4}\right) \\
 &\left.\times \log \left(1-\frac{z}{3}\right)\right],
 \\
\notag   d_{6,6}\left(g \rightarrow 0^{++}\right)=& \frac{\pi^{2} \alpha_{s}^{4}}{55296 z(2-z)^{2}(3-z)}\left[186624-430272 z+617824 z^{2}-634902 z^{3}\right.\\
\notag  &+374489 z^{4}-115387 z^{5}+14378 z^{6}-6(1-z)(2-z)(3-z)(-144-2166 z\\
\notag  &\left.+1015 z^{2}+70 z^{3}\right) \log (1-z)-156(2-z)(3-z)\left(144-1242 z+1693 z^{2}-876 z^{3}\right.\\
\notag  &\left.+170 z^{4}\right) \log \left(1-\frac{z}{2}\right)+300(2-z)(3-z)\left(72-714 z+953 z^{2}-472 z^{3}+87 z^{4}\right) \\
 &\left.\times \log \left(1-\frac{z}{3}\right)\right],
 \\
\notag  d_{3,6}\left(g \rightarrow 0^{++}\right)=& \frac{\pi^{2} \alpha_{s}^{4}}{165888 z(2-z)^{2}(3-z)}\left[186624-430272 z+490720 z^{2}-394422 z^{3}\right.\\
\notag  &+199529 z^{4}-57547 z^{5}+7082 z^{6}+6(1-z)(2-z)(3-z)(-432+3302 z\\
\notag  &\left.-1855 z^{2}+210 z^{3}\right) \log (1-z)-12(2-z)(3-z)\left(720-2258 z+2329 z^{2}-1052 z^{3}\right.\\
\notag  &\left.+226 z^{4}\right) \log \left(1-\frac{z}{2}\right)+12(2-z)(3-z)\left(936-4882 z+4989 z^{2}-1936 z^{3}+331 z^{4}\right) \\
 &\left.\times \log \left(1-\frac{z}{3}\right)\right].
\end{align}
\label{dz:0++}
\eseq
For gluon fragmentation into $2^{++}$ tetraquark, the three SDCs listed
in \eqref{NRQCD:fac:Tc:fragmentation} read
\begin{subequations}
\begin{align}
\notag  d_{3,3}\left(g \rightarrow 2^{++}\right)=& \frac{\pi^{2} \alpha_{s}^{4}}{622080 z^{2}(2-z)^{2}(3-z)}\left[2\left(46656-490536 z+1162552 z^{2}-1156308 z^{3}\right.\right.\\
\notag  &\left.+595421 z^{4}-170578 z^{5}+21212 z^{6}\right) z+3(1-z)(2-z)(3-z)(-20304-31788 z)(1296+1044 z\\
 &\left.\left.+73036 z^{2}-36574 z^{3}+7975 z^{4}\right) \log (1-z)+33(2-z)\notag (3-z)(1296+25)\right] \\
  &\left.\left.-9224 z^{2}+9598 z^{3}-3943 z^{4}+725 z^{5}\right) \log \left(1-\frac{z}{3}\right)\right],\\
 d_{6,6}\left(g \rightarrow 2^{++}\right)=& d_{3,6}\left(g \rightarrow 2^{++}\right)=0.
\end{align}
\label{dz:2++}
\end{subequations}

The limit behavior of various SDCs is particularly simple, exhibiting $1/z$ scaling:
\begin{subequations}
\bqa
    && d_{3,3}(g\to 0^{++})\to\frac{\pi ^2 \alpha _s^4}{32 z},
    \qquad d_{6,6}(g\to 0^{++})\to\frac{9\pi ^2 \alpha _s^4}{32 z},\qquad  d_{3,6}(g\to 0^{++})\to\frac{3\pi ^2 \alpha _s^4}{32 z},
\\
    && d_{3,3}(g\to 2^{++})\to\frac{\pi ^2 \alpha _s^4}{20 z}.
\eqa
\end{subequations}

\paragraph{\color{blue} Phenomenology on $T_{4c}$ production at LHC\label{res}\;}

To predict $T_{4c}$ fragmentation production at {\tt LHC}, we need some definite knowledge about three
nonperturbative NRQCD matrix elements in \eqref{NRQCD:fac:Tc:fragmentation}. In principle these LDMEs are amenable to
future lattice NRQCD simulation. To proceed, we resort to the popular diquark model, {\it i.e.}, the $T_{4c}$ is bound by
the diquark and antiquark cluster by some phenomenological potential. To make life simpler, we neglect the Fock component
of $\mathbf{6}\otimes\mathbf{\bar 6}$ and only retain $\mathbf{3}\otimes\mathbf{\bar 3}$~\footnote{We note this phenomenological
model may be an oversimplification, not fully justified from the first principle of QCD. Rigorously speaking,
$T_{4c}$ should be a compact tetraquark where there is no scale hierarchy between diquark size and diquark-antidiquark separation~\cite{Chao:2020dml}.}.
In this context, we may identify the NRQCD LDME with the product of phenomenological wave functions at the origin:
\bseq
\begin{align}
& \left\langle 0\left|\mathcal{O}^{(0)}\right|0^{++} \right\rangle
\approx \frac{1}{\pi^{3 / 2}} R_{\mathcal{D}}^{2}(0) R_{T}(0),
\\
& \left\langle 0 \left|{\mathcal O}^{(2)}_{\alpha \beta} \right| 2^{++}, m_j \right\rangle \approx
\frac{\varepsilon_{\alpha \beta}^{\left(m_{j}\right)}}{\pi^{3/2}} R_{\mathcal{D}}^{2}(0) R_{T}(0),
\end{align}
\eseq
where $m_j$ denotes the magnetic number of the $2^{++}$ tetraquark and
$\varepsilon_{\alpha\beta}^{(m_j)}$ is its polarization tensor. $R_D(0)$ denotes the
wave function at the origin for the diquark system, while $R_T(0)$ represents the
wave function at the origin for the diquark-antidiquark cluster.

Squaring the vacuum-to-$T_{4c}$ matrix element and summing over the polarization,
of the LDMEs,  taking into account the symmetry factor of $1/4$ arising from Fermi statistics,
we end up with the intended NRQCD LDMEs appearing in the factorization formula \eqref{NRQCD:fac:Tc:fragmentation}:
\bseq
\begin{align}
&\left|{\left\langle 0\left|{\mathcal{O}^{(0)}}\right|{T_{4c}^{(0)}}\right\rangle}\right|^2=\frac{1}{4\pi^3}
\left|{R_\mathcal{D}(0)}\right|^4\left|{R_T(0)}\right|^2,
\\
  \sum_{m_j}&\left|{\left\langle 0\left|{\mathcal{O}_{\alpha \beta}^{(2)}}\right|{T_{4c}^{(2,m_j)}}\right\rangle}\right|^2=\frac{5}{4\pi^3}\left|{R_\mathcal{D}(0)}\right|^4\left|{R_T(0)}\right|^2.
\end{align}
\eseq
The occurrence of the factor 5 reflects the approximate heavy diquark spin symmetry.

In phenomenological analysis for $T_{4c}$, we choose $R_{\mathcal{D}}(0)=0.523\;\mathrm{GeV}^{3/2}$~\cite{Kiselev:2002iy}.
The wave function at the origin for diquark-antidiquark cluster is computed in \cite{Debastiani:2017msn,Berezhnoy:2011xy,Berezhnoy:2012bv}.
We take $R_T(0)=2.902 \;\mathrm{GeV}^{3/2}$ resulting from Cornell-type potential model~\cite{Debastiani:2017msn}.
For $T_{4b}$, assuming both the inter-quark and inter-diquark potentials are Coulombic, we then find
$R_{\mathcal{D}_b}^{\mathrm{Coul}}(0)=0.703\;\mathrm{GeV}^{3/2}$ and $R_{T_b}^{\mathrm{Coul}}(0)=5.579 \;\mathrm{GeV}^{3/2}$.

\begin{figure}[!hbtp]
\centering
\includegraphics[width=0.95\textwidth]{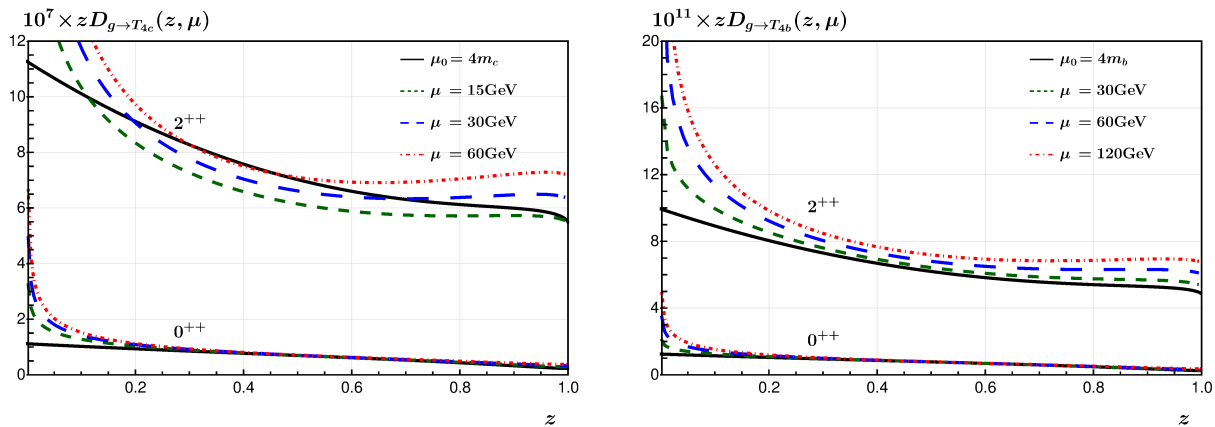}
\caption{Gluon-to-$T_{4c/4b}$ fragmentation functions at various scales.
The black solid curves represent the fragmentation functions at initial scales.}
\label{fig:zgFFevl}
\end{figure}

We take $m_c=1.5$ GeV and $m_b=4.8$ GeV. The fragmentation functions $D_{g\to T_{4c,4b}}(z,\mu)$ in \eqref{NRQCD:fac:Tc:fragmentation},
\eqref{dz:0++} and \eqref{dz:0++} are given at the default initial scale $\mu_0=4m_Q$.
We then evolve them to higher scales in accordance with the DGLAP equation \eqref{DGLAP:evolu:eq}, only
keeping the gluon-gluon splitting kernel. We employ the 4-order Runge-Kutta method with the assistance of {\tt{GNU Scientific Library}}~\cite{GSL}, and illustrate the evolution effect in Fig.~\ref{fig:zgFFevl}.

\begin{figure}[!hbtp]
	\centering
	\includegraphics[width=0.85\textwidth]{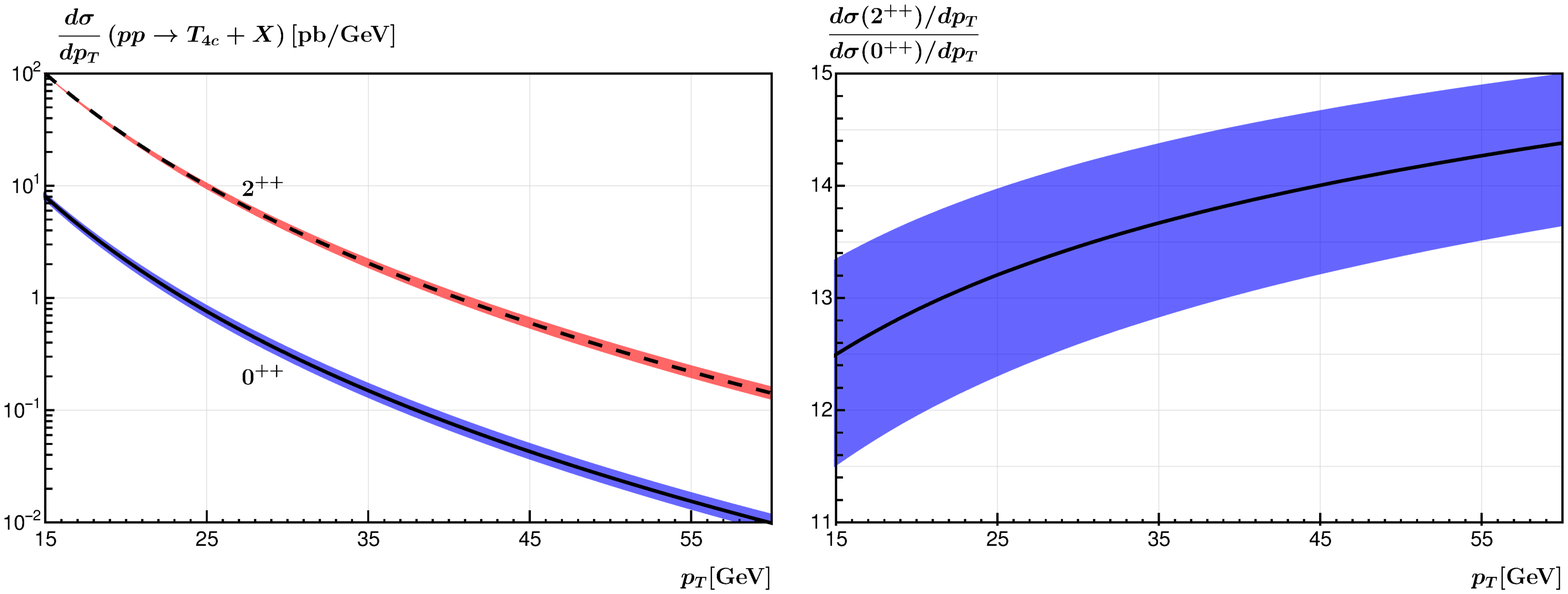}
	\includegraphics[width=0.85\textwidth]{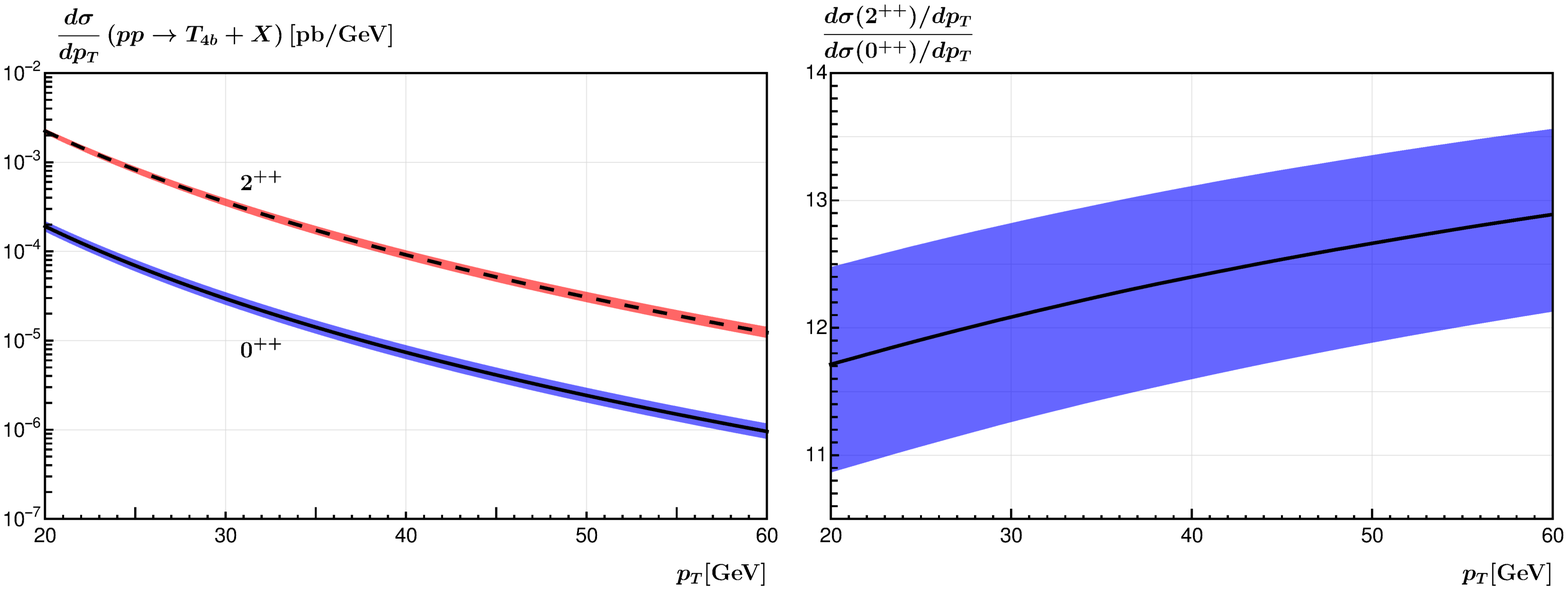}
	\caption{The $p_T$ distribution of inclusive $T_{4c/4b}$ production on LHC. The central values (represented by the solid and dashed curves) are generated by setting $\mu=p_T$. The difference between $0^{++}$ and $2^{++}$ states are also given. }
	\label{fig:pT_dstrbtn}
\end{figure}

We then compute the fully heavy tetraquark $p_T$ spectrum in $pp$ collision
with $\sqrt{s}=13\mathrm{TeV}$, according to the QCD factorization theorem \eqref{QCD:fact:theor}.
We utilize the {\tt CTEQ14} sets~\cite{Dulat:2015mca} for gluon PDF,
and set both the factorization scales appearing in PDF and fragmentation function
to be equal. We also impose the rapidity cut $-5\leq y \leq 5$.
In order to estimate the theoretical error, we slide the factorization scale $\mu$ by from $p_T/2$ to $2p_T$, with the central value
$p_T$. We remark the quark mass uncertainty may lead to much large error.
The numerical prediction for the $p_T$ spectrum of the $0^{++}$ and $2^{++}$ $T_{4c/4b}$ are shown in Fig.~\ref{fig:pT_dstrbtn}.
The integrated cross sections are also tabulated in Table.~\ref{pT_intgrtd}.
\begin{table}[h]
\begin{tabular}{|c|c|c|c|c|c|}
    \hline
     &  & \multicolumn{2}{c|}{$0^{++}$} & \multicolumn{2}{c|}{$2^{++}$ }\tabularnewline
    \hline
     & $p_{T}$ range & $\sigma$  & $N_{\mathrm{events}}$ & $\sigma$  & $N_{\mathrm{events}}$\tabularnewline
    \hline
    $T_{4c}$ & $15\mathrm{\mathrm{GeV}}\leq p_{T}\leq60\mathrm{\mathrm{GeV}}$  & $33_{-4}^{+4}\mathrm{pb}$ & $9.9_{-1.2}^{+1.2}\times10^{7}$ & $424_{-21}^{+13}\mathrm{pb}$ & $1.27_{-0.06}^{+0.04}\times10^{9}$\tabularnewline
    \hline
    $T_{4b}$ & $20\mathrm{\mathrm{GeV}}\leq p_{T}\leq60\mathrm{\mathrm{GeV}}$  & $1.04_{-0.15}^{+0.17}\times10^{-3}\mathrm{pb}$ & $3.12_{-0.45}^{+0.51}\times10^{3}$ & $1.24_{-0.11}^{+0.11}\times10^{-2}\mathrm{pb}$ & $3.72_{-0.33}^{+0.33}\times10^{4}$\tabularnewline
    \hline
    \end{tabular}
\caption{The $p_T$-integrated cross section for $T_4c$ inclusive production on LHC.}
\label{pT_intgrtd}
\end{table}

It is interesting to note that the $p_T$ spectrum of $2^{++}$ sate is about an order of magnitude greater than that of $0^{++}$ sate. We present both $T_{4c}$'s and $T_{4b}$'s $p_T$-spectrum for comparison, the former is about $10^4$ larger than the latter.

There are other estimations of the cross sections: Karliner et al. roughly estimated the cross section of $T_{4c}$ production to be 0.7 nb, and $T_{4b}$ to be 1 pb at $\sqrt{s}=7$ TeV\cite{Karliner:2016zzc}; Carvalho et al. arrived with $3.6\pm2.5$ nb at $\sqrt{s}=7$ TeV and $7.0\pm4.8$ nb at $\sqrt{s}=14$ TeV\cite{Carvalho:2015nqf} for $T_{4c}$, and $13.9\pm10.1$ pb at $\sqrt{s}=14$ TeV for $T_{4b}$\cite{Carvalho:2016shs}. Despite the obvious discrepancy between our result and theirs, our integrated cross section, limited by the fact that fragmentation is only valid at large-$p_T$, does not account for the small-$p_T$ region.

Based on the cross sections obtained in Table I, and the integrated luminosity 3000 $fb^{-1}$, we obtain the yields of the accumulated event number for $T_{4c}$ at LHC are a hundred million for $0^{++}$ and 8 hundreds million for $2^{++}$. It is observed that the
prediction for $T_{4b}$ is highly suppressed, mainly due to the relative larger bottom mass suppression.
It seems that the cross section appeared in Table I is apparently smaller than the value in Ref.~\cite{Karliner:2016zzc,Carvalho:2015nqf,Carvalho:2016shs}.
It is noted that the cross section of tetraquark in our calculation concentrates on the low $p_T$ region, thus its value is sensitive to the $p_T$ cut.
If we extrapolate the selection $p_T\geq 5.2 GeV$, the cross section for $T_{4c}$ may reach several $nb$, which is compatible with the
phenomenology prediction in Ref.~\cite{Karliner:2016zzc,Carvalho:2015nqf,Carvalho:2016shs}. To the contrary, our predicted cross section for $T_{4b}$ is several order-of-magnitude smaller than the value in Ref.~\cite{Karliner:2016zzc,Carvalho:2015nqf,Carvalho:2016shs}, where the double parton scattering is considered.

\paragraph{\color{blue}Summary\label{summary}}
The recent discovery of $X(6900)$ resonance has renewed the interest toward the tetraquark composed of fully heavy quarks.
Although much has to be learned about the binding mechanism for this novel type of tetraquarks, in this work we propose
a model-independent approach to study the inclusive production of fully heavy tetraquark at large $p_T$, which is
based on NRQCD factorization of the fragmentation function, owing to the asymptotic freedom of QCD.
Due to the huge luminosity at {\tt LHC}, the production rate of $T_{4c}$ appears to be significant, and
we hope that the future precise measurements of the $p_T$ spectrum can shed important on 
the production mechanism of the fully heavy tetraquarks.

\paragraph{\color{blue}Acknowledgments}
\begin{acknowledgments}
We thank Dingyu Shao for useful discussions.	
The work of F.~F. is supported by the National Natural
Science Foundation of China under Grant No. 11875318,
No. 11505285, and by the Yue Qi Young Scholar Project
in CUMTB.
The work of Y.-S.~H., Y.~J. and J.-Y.~Z. is supported in part by the National Natural Science Foundation of China under Grants No.~11925506, 11875263, No.~11621131001 (CRC110 by DFG and NSFC).
The work of W.-L. S. is supported by the National Natural Science Foundation
of China under Grants No. 11975187 and the Natural Science Foundation of ChongQing under Grant No. cstc2019jcyj-msxm2667.
The work of X.-N. X. is supported in part by the National Natural Science Foundation of China under Grants No. 11905296.
\end{acknowledgments}

\appendix


\begin{thebibliography}{100}

  \bibitem{Liu:2019zoy}
  Y.~R.~Liu, H.~X.~Chen, W.~Chen, X.~Liu and S.~L.~Zhu,
    Prog.\ Part.\ Nucl.\ Phys.\  {\bf 107}, 237 (2019)
    doi:10.1016/j.ppnp.2019.04.003
    [arXiv:1903.11976 [hep-ph]].

  \bibitem{Chen:2016qju}
  H.~X.~Chen, W.~Chen, X.~Liu and S.~L.~Zhu,
    Phys.\ Rept.\  {\bf 639}, 1 (2016)
    doi:10.1016/j.physrep.2016.05.004
    [arXiv:1601.02092 [hep-ph]].

  \bibitem{Klempt:2007cp}
  E.~Klempt and A.~Zaitsev,
    Phys.\ Rept.\  {\bf 454}, 1 (2007)
    doi:10.1016/j.physrep.2007.07.006
    [arXiv:0708.4016 [hep-ph]].

  \bibitem{Hosaka:2016pey}
  A.~Hosaka, T.~Iijima, K.~Miyabayashi, Y.~Sakai and S.~Yasui,
    PTEP {\bf 2016}, no. 6, 062C01 (2016)
    doi:10.1093/ptep/ptw045
    [arXiv:1603.09229 [hep-ph]].

  \bibitem{Ali:2017jda}
  A.~Ali, J.~S.~Lange and S.~Stone,
    Prog.\ Part.\ Nucl.\ Phys.\  {\bf 97}, 123 (2017)
    doi:10.1016/j.ppnp.2017.08.003
    [arXiv:1706.00610 [hep-ph]].

  \bibitem{Aaij:2020fnh}
  R.~Aaij {\it et al.} [LHCb Collaboration],
    arXiv:2006.16957 [hep-ex].

  \bibitem{Iwasaki:1976cn}
  Y.~Iwasaki,
    Phys.\ Rev.\ Lett.\  {\bf 36}, 1266 (1976).
    doi:10.1103/PhysRevLett.36.1266

  \bibitem{Chao:1980dv}
  K.~T.~Chao,
    Z.\ Phys.\ C {\bf 7}, 317 (1981).
    doi:10.1007/BF01431564

  \bibitem{Ader:1981db}
  J.~P.~Ader, J.~M.~Richard and P.~Taxil,
    Phys.\ Rev.\ D {\bf 25}, 2370 (1982).
    doi:10.1103/PhysRevD.25.2370

  \bibitem{Wu:2016vtq}
  J.~Wu, Y.~R.~Liu, K.~Chen, X.~Liu and S.~L.~Zhu,
    Phys.\ Rev.\ D {\bf 97}, no. 9, 094015 (2018)
    doi:10.1103/PhysRevD.97.094015
    [arXiv:1605.01134 [hep-ph]].

  \bibitem{Barnea:2006sd}
  N.~Barnea, J.~Vijande and A.~Valcarce,
    Phys.\ Rev.\ D {\bf 73}, 054004 (2006)
    doi:10.1103/PhysRevD.73.054004
    [hep-ph/0604010].

  \bibitem{Badalian:1985es}
  A.~M.~Badalian, B.~L.~Ioffe and A.~V.~Smilga,
    Nucl.\ Phys.\ B {\bf 281}, 85 (1987).
    doi:10.1016/0550-3213(87)90248-3

  \bibitem{Liu:2019zuc}
  M.~S.~Liu, Q.~F.~Lü, X.~H.~Zhong and Q.~Zhao,
    Phys.\ Rev.\ D {\bf 100}, no. 1, 016006 (2019)
    doi:10.1103/PhysRevD.100.016006
    [arXiv:1901.02564 [hep-ph]].

  \bibitem{Bedolla:2019zwg}
  M.~A.~Bedolla, J.~Ferretti, C.~D.~Roberts and E.~Santopinto,
    arXiv:1911.00960 [hep-ph].

  \bibitem{Lloyd:2003yc}
  R.~J.~Lloyd and J.~P.~Vary,
    Phys.\ Rev.\ D {\bf 70}, 014009 (2004)
    doi:10.1103/PhysRevD.70.014009
    [hep-ph/0311179].

  \bibitem{Anwar:2017toa}
  M.~N.~Anwar, J.~Ferretti, F.~K.~Guo, E.~Santopinto and B.~S.~Zou,
    Eur.\ Phys.\ J.\ C {\bf 78}, no. 8, 647 (2018)
    doi:10.1140/epjc/s10052-018-6073-9
    [arXiv:1710.02540 [hep-ph]].

  \bibitem{Esposito:2018cwh}
  A.~Esposito and A.~D.~Polosa,
    Eur.\ Phys.\ J.\ C {\bf 78}, no. 9, 782 (2018)
    doi:10.1140/epjc/s10052-018-6269-z
    [arXiv:1807.06040 [hep-ph]].

  \bibitem{Bai:2016int}
  Y.~Bai, S.~Lu and J.~Osborne,
    Phys.\ Lett.\ B {\bf 798}, 134930 (2019)
    doi:10.1016/j.physletb.2019.134930
    [arXiv:1612.00012 [hep-ph]].

  \bibitem{Richard:2017vry}
  J.~M.~Richard, A.~Valcarce and J.~Vijande,
    Phys.\ Rev.\ D {\bf 95}, no. 5, 054019 (2017)
    doi:10.1103/PhysRevD.95.054019
    [arXiv:1703.00783 [hep-ph]].

  \bibitem{Berezhnoy:2011xn}
  A.~V.~Berezhnoy, A.~V.~Luchinsky and A.~A.~Novoselov,
    Phys.\ Rev.\ D {\bf 86}, 034004 (2012)
    doi:10.1103/PhysRevD.86.034004
    [arXiv:1111.1867 [hep-ph]].

  \bibitem{Becchi:2020uvq}
  C.~Becchi, A.~Giachino, L.~Maiani and E.~Santopinto,
    arXiv:2006.14388 [hep-ph].

  \bibitem{Heller:1985cb}
  L.~Heller and J.~A.~Tjon,
    Phys.\ Rev.\ D {\bf 32}, 755 (1985).
    doi:10.1103/PhysRevD.32.755

  \bibitem{Debastiani:2017msn}
  V.~R.~Debastiani and F.~S.~Navarra,
    Chin.\ Phys.\ C {\bf 43}, no. 1, 013105 (2019)
    doi:10.1088/1674-1137/43/1/013105
    [arXiv:1706.07553 [hep-ph]].

  \bibitem{Lu:2020cns}
  Q.~F.~Lü, D.~Y.~Chen and Y.~B.~Dong,
    arXiv:2006.14445 [hep-ph].

  \bibitem{liu:2020eha}
  M.~S.~liu, F.~X.~Liu, X.~H.~Zhong and Q.~Zhao,
    arXiv:2006.11952 [hep-ph].

  \bibitem{Yang:2020rih}
  G.~Yang, J.~Ping, L.~He and Q.~Wang,
    arXiv:2006.13756 [hep-ph].

  \bibitem{Chen:2016jxd}
  W.~Chen, H.~X.~Chen, X.~Liu, T.~G.~Steele and S.~L.~Zhu,
    Phys.\ Lett.\ B {\bf 773}, 247 (2017)
    doi:10.1016/j.physletb.2017.08.034
    [arXiv:1605.01647 [hep-ph]].

  \bibitem{Chen:2018cqz}
  W.~Chen, H.~X.~Chen, X.~Liu, T.~G.~Steele and S.~L.~Zhu,
    EPJ Web Conf.\  {\bf 182}, 02028 (2018)
    doi:10.1051/epjconf/201818202028
    [arXiv:1803.02522 [hep-ph]].

  \bibitem{Wang:2017jtz}
  Z.~G.~Wang,
    Eur.\ Phys.\ J.\ C {\bf 77}, no. 7, 432 (2017)
    doi:10.1140/epjc/s10052-017-4997-0
    [arXiv:1701.04285 [hep-ph]].

  \bibitem{Wang:2018poa}
  Z.~G.~Wang and Z.~Y.~Di,
    Acta Phys.\ Polon.\ B {\bf 50}, 1335 (2019)
    doi:10.5506/APhysPolB.50.1335
    [arXiv:1807.08520 [hep-ph]].

  \bibitem{Chen:2020xwe}
  H.~X.~Chen, W.~Chen, X.~Liu and S.~L.~Zhu,
    arXiv:2006.16027 [hep-ph].

  \bibitem{Wang:2020ols}
  Z.~G.~Wang,
    arXiv:2006.13028 [hep-ph].
    
 \bibitem{Karliner:2020dta} M.~Karliner and J.~L.~Rosner, 
 [arXiv:2009.04429 [hep-ph]]. 

  \bibitem{Hughes:2017xie}
  C.~Hughes, E.~Eichten and C.~T.~H.~Davies,
    Phys.\ Rev.\ D {\bf 97}, no. 5, 054505 (2018)
    doi:10.1103/PhysRevD.97.054505
    [arXiv:1710.03236 [hep-lat]].

  \bibitem{Berezhnoy:2011xy}
  A.~V.~Berezhnoy, A.~K.~Likhoded, A.~V.~Luchinsky and A.~A.~Novoselov,
    Phys.\ Rev.\ D {\bf 84}, 094023 (2011)
    doi:10.1103/PhysRevD.84.094023
    [arXiv:1101.5881 [hep-ph]].

  \bibitem{Karliner:2016zzc}
  M.~Karliner, S.~Nussinov and J.~L.~Rosner,
    Phys.\ Rev.\ D {\bf 95}, no. 3, 034011 (2017)
    doi:10.1103/PhysRevD.95.034011
    [arXiv:1611.00348 [hep-ph]].

  \bibitem{Becchi:2020mjz}
  C.~Becchi, A.~Giachino, L.~Maiani and E.~Santopinto,
    Phys.\ Lett.\ B {\bf 806}, 135495 (2020)
    doi:10.1016/j.physletb.2020.135495
    [arXiv:2002.11077 [hep-ph]].

  \bibitem{Carvalho:2015nqf}
  F.~Carvalho, E.~R.~Cazaroto, V.~P.~Gonçalves and F.~S.~Navarra,
    Phys.\ Rev.\ D {\bf 93}, no. 3, 034004 (2016)
    [Phys.\ Rev.\ D {\bf 93}, 034004 (2016)]
    doi:10.1103/PhysRevD.93.034004
    [arXiv:1511.05209 [hep-ph]].

  \bibitem{Maciula:2020wri}
  R.~Maciuła, W.~Schäfer and A.~Szczurek,
    arXiv:2009.02100 [hep-ph].

  \bibitem{Jin:2020jfc}
  X.~Jin, Y.~Xue, H.~Huang and J.~Ping,
    arXiv:2006.13745 [hep-ph].

  \bibitem{Chao:2020dml}
  K.~T.~Chao and S.~L.~Zhu,
  doi:10.1016/j.scib.2020.08.031
  [arXiv:2008.07670 [hep-ph]].

  \bibitem{Collins:1989gx}
  J.~C.~Collins, D.~E.~Soper and G.~F.~Sterman,
  Adv. Ser. Direct. High Energy Phys. {\bf 5}, 1-91 (1989)
  doi:10.1142/9789814503266\_0001
  [arXiv:hep-ph/0409313 [hep-ph]].

  \bibitem{Collins:1981uw}
  J.~C.~Collins and D.~E.~Soper,
    Nucl.\ Phys.\ B {\bf 194}, 445 (1982).
    doi:10.1016/0550-3213(82)90021-9

  \bibitem{Braaten:1994xb}
  E.~Braaten, M.~A.~Doncheski, S.~Fleming and M.~L.~Mangano,
    Phys.\ Lett.\ B {\bf 333}, 548 (1994)
    doi:10.1016/0370-2693(94)90182-1
    [hep-ph/9405407].

  \bibitem{Ma:1994zt}
  J.~P.~Ma,
    Phys.\ Lett.\ B {\bf 332}, 398 (1994)
    doi:10.1016/0370-2693(94)91271-8
    [hep-ph/9401249].

\bibitem{Ma:2013yla}
  Y.~Q.~Ma, J.~W.~Qiu and H.~Zhang,
  Phys.\ Rev.\ D {\bf 89}, no. 9, 094029 (2014)
  doi:10.1103/PhysRevD.89.094029
  [arXiv:1311.7078 [hep-ph]].

  \bibitem{Nogueira:1991ex}
  P.~Nogueira,
    J.\ Comput.\ Phys.\  {\bf 105}, 279 (1993).
    doi:10.1006/jcph.1993.1074

  \bibitem{Bauer:2000cp}
  C.~W.~Bauer, A.~Frink and R.~Kreckel,
    J.\ Symb.\ Comput.\  {\bf 33}, 1 (2002)
    doi:10.1006/jsco.2001.0494
    [cs/0004015 [cs-sc]].

  \bibitem{Petrelli:1997ge}
  A.~Petrelli, M.~Cacciari, M.~Greco, F.~Maltoni and M.~L.~Mangano,
  Nucl. Phys. B \textbf{514}, 245-309 (1998)
  doi:10.1016/S0550-3213(97)00801-8
  [arXiv:hep-ph/9707223 [hep-ph]].

  \bibitem{Braaten:2002fi}
  E.~Braaten and J.~Lee,
  Phys. Rev. D \textbf{67}, 054007 (2003)
  doi:10.1103/PhysRevD.72.099901
  [arXiv:hep-ph/0211085 [hep-ph]].

  \bibitem{Hahn:2000kx}
  T.~Hahn,
    Comput.\ Phys.\ Commun.\  {\bf 140}, 418 (2001)
    doi:10.1016/S0010-4655(01)00290-9
    [hep-ph/0012260].

  \bibitem{Shtabovenko:2016sxi}
  V.~Shtabovenko, R.~Mertig and F.~Orellana,
    Comput.\ Phys.\ Commun.\  {\bf 207}, 432 (2016)
    doi:10.1016/j.cpc.2016.06.008
    [arXiv:1601.01167 [hep-ph]].

  \bibitem{Kiselev:2002iy}
  V.~V.~Kiselev, A.~K.~Likhoded, O.~N.~Pakhomova and V.~A.~Saleev,
    Phys.\ Rev.\ D {\bf 66}, 034030 (2002)
    doi:10.1103/PhysRevD.66.034030
    [hep-ph/0206140].


  \bibitem{Berezhnoy:2012bv}
  A.~V.~Berezhnoy, A.~K.~Likhoded, A.~V.~Luchinsky and A.~A.~Novoselov,
    Phys.\ Atom.\ Nucl.\  {\bf 75}, 1006 (2012)
    [Yad.\ Fiz.\  {\bf 75}, 1067 (2012)].
    doi:10.1134/S1063778812040035

  \bibitem{Isgur:1989vq}
  N.~Isgur and M.~B.~Wise,
    Phys.\ Lett.\ B {\bf 232}, 113 (1989).
    doi:10.1016/0370-2693(89)90566-2

  \bibitem{GSL}
  Galassi, M., Davies, J., Theiler, J., Gough, B., Jungman, G., Alken, P., Booth, M., Rossi, F. and Ulerich, R., 2002.
  GNU scientific library. Network Theory Limited.

\bibitem{Dulat:2015mca} S.~Dulat, T.~J.~Hou, J.~Gao, M.~Guzzi, J.~Huston, P.~Nadolsky, J.~Pumplin, C.~Schmidt, D.~Stump and C.~P.~Yuan,
Phys. Rev. D \textbf{93}, no.3, 033006 (2016)
doi:10.1103/PhysRevD.93.033006 [arXiv:1506.07443 [hep-ph]].


\bibitem{Carvalho:2016shs}
E.~R.~Cazaroto, F.~Carvalho, V.~P.~Gonçalves and F.~S.~Navarra,
PoS \textbf{LHCP2016}, 178 (2016)
10.22323/1.276.0178

\end{thebibliography}

\end{document}